# Oceanic and atmospheric methane cycling in the cGENIE Earth system model

Christopher T. Reinhard[1,2,3*], Stephanie L. Olson[2,4,5], Sandra Kirtland Turner[6], Cecily Pälike[7], Yoshiki Kanzaki[6], Andy Ridgwell[6]

[1]School of Earth and Atmospheric Sciences, Georgia Institute of Technology, Atlanta, GA 30332
[2]NASA Astrobiology Institute, Alternative Earths Team, Riverside, CA
[3]NASA Nexus for Exoplanet System Science (NExSS) Upside-Down Biospheres Team, Georgia Institute of Technology, Atlanta, GA
[4]Department of Geophysical Sciences, University of Chicago, Chicago, IL 60637
[5]Department of Earth, Atmospheric, and Planetary Science, Purdue University, West Lafayette, IN 47907
[6]Department of Earth Sciences, University of California, Riverside, Riverside, CA 92521
[7]MARUM Center for Marine Environmental Sciences, University of Bremen, Germany

*To whom correspondence should be addressed. E-mail: chris.reinhard@eas.gatech.edu

**Abstract:** The methane ($CH_4$) cycle is a key component of the Earth system that links planetary climate, biological metabolism, and the global biogeochemical cycles of carbon, oxygen, sulfur, and hydrogen. However, currently lacking is a numerical model capable of simulating a diversity of environments in the ocean where $CH_4$ can be produced and destroyed, and with the flexibility to be able to explore not only relatively recent perturbations to Earth's $CH_4$ cycle but also to probe $CH_4$ cycling and associated climate impacts under the very low-$O_2$ conditions characteristic of most of Earth history and likely widespread on other Earth-like planets. Here, we present a refinement and expansion of the ocean-atmosphere $CH_4$ cycle in the intermediate-complexity Earth system model cGENIE, including parameterized atmospheric $O_2$-$O_3$-$CH_4$ photochemistry and schemes for microbial methanogenesis, aerobic methanotrophy, and anaerobic oxidation of methane (AOM). We describe the model framework, compare model parameterizations against modern observations, and illustrate the flexibility of the model through a series of example simulations. Though we make no attempt to rigorously tune default model parameters, we find that simulated atmospheric $CH_4$ levels and marine dissolved $CH_4$ distributions are generally in good agreement with empirical constraints for the modern and recent Earth. Finally, we illustrate the model's utility in understanding the time-dependent behavior of the $CH_4$ cycle resulting from transient carbon injection into the atmosphere, and present model ensembles that examine the effects of atmospheric $pO_2$, oceanic dissolved $SO_4^{2-}$, and the thermodynamics of microbial metabolism on steady-state atmospheric $CH_4$ abundance. Future model developments will address the sources and sinks of $CH_4$ associated with the terrestrial biosphere and marine $CH_4$ gas hydrates, both of which will be essential for comprehensive treatment of Earth's $CH_4$ cycle during geologically recent time periods.

## 1. Introduction

The global biogeochemical cycle of methane ($CH_4$) is central to the evolution and climatic stability of the Earth system. Methane provides an important substrate for microbial metabolism, particularly in energy-limited microbial ecosystems in the deep subsurface (Valentine, 2011;Chapelle et al., 1995) and in anoxic marine and lacustrine sediments (Lovley et al., 1982;Hoehler et al., 2001). Indeed, the microbial production and consumption of $CH_4$ are amongst the oldest metabolisms on Earth, with an isotopic record of bacterial methane cycling stretching back nearly 3.5 billion years (Ueno et al., 2006;Hinrichs, 2002;Hayes, 1994). As the most abundant hydrocarbon in Earth's atmosphere $CH_4$ also has a significant influence on atmospheric photochemistry (Thompson and Cicerone, 1986), and because it absorbs in a window region of Earth's longwave emission spectrum it is an important greenhouse gas. This has important implications over the coming centuries, with atmospheric $CH_4$ classified as a critical near-term climate forcing (Myhre et al., 2013), but has also resulted in dramatic impacts during certain periods of Earth history. For example, high steady-state atmospheric $CH_4$ has been invoked as an important component of Earth's early energy budget, potentially helping to offset a dim early Sun (Sagan and Mullen, 1972;Pavlov et al., 2000;Haqq-Misra et al., 2008), while time-dependent changes to the atmospheric $CH_4$ inventory have been invoked as drivers of extreme climatic perturbations throughout Earth history (Dickens et al., 1997;Dickens, 2003;Bjerrum and Canfield, 2011;Zeebe, 2013;Schrag et al., 2002). Because it is cycled largely through biological processes on the modern (and ancient) Earth and is spectrally active, atmospheric $CH_4$ has also been suggested as a remotely detectable biosignature that could be applied to planets beyond our solar system (Hitchcock and Lovelock, 1967;Sagan et al., 1993;Krissansen-Totton et al., 2018).

A number of low-order Earth system models incorporating a basic $CH_4$ cycle have been developed, particularly with a view to addressing relatively 'deep time' geological questions. These include explorations of long-term changes to the chemistry of Earth's atmosphere (Claire et al., 2006;Catling et al., 2007;Bartdorff et al., 2008;Beerling et al., 2009), potential climate impacts at steady state (Kasting et al., 2001;Ozaki et al., 2018), and transient impacts of $CH_4$ degassing on climate (Schrag et al., 2002;Bjerrum and Canfield, 2011). In some cases these models explicitly couple surface fluxes to a model of atmospheric photochemistry (Lamarque et al., 2006;Ozaki et al., 2018;Kasting et al., 2001), but in general atmospheric chemistry is parameterized based on offline 1- or 2-D photochemical models while surface fluxes are specified arbitrarily or are based on a simple 1-box ocean-biosphere model. A range of slightly more complex 'box' model approaches have been applied to simulate transient perturbations to Earth's $CH_4$ cycle and attendant climate impacts on timescales ranging from ~$10^5$ years (Dickens et al., 1997;Dickens, 2003) to ~$10^8$ years (Daines and Lenton, 2016). In addition, offline and/or highly parameterized approaches toward simulating the impact of transient $CH_4$ degassing from gas hydrate reservoirs have been developed and applied to relatively recent periods of Earth history (Archer and Buffett, 2005;Lunt et al., 2011) or projected future changes (Archer et al., 2009;Hunter et al., 2013). However, the most sophisticated and mechanistic models of global $CH_4$ cycling currently available tend to focus on terrestrial (soil or wetland) sources and sinks (Ridgwell et al., 1999;Walter and Heimann, 2000;Wania et al., 2010;Konijnendijk et al., 2011;Melton et al., 2013) or focus on explicitly modeling atmospheric photochemistry (Shindell et al., 2013).

Much less work has been done to develop ocean biogeochemistry models that are both equipped to deal with the wide range of boundary conditions characteristic of Earth history and are computationally tractable when running large model ensembles and/or on long (approaching ~$10^6$ year) timescales, as well as being able to simulate the (3-D) redox structure of the ocean allowing for localized zones of production and oxidation (which provides more accurate estimates of emission to the atmosphere). For instance, Elliot et al. (2011) advanced modelling of marine $CH_4$ cycling by developing and employing a 3-D ocean circulation and climate model (CCSM-3) to simulate the impact of injecting clathrate-derived $CH_4$ into the Arctic ocean. However, microbial consumption of $CH_4$ in the ocean interior was parameterized via an empirical log-linear function that implicitly neglects anaerobic oxidation of methane (AOM) via dissolved sulfate ($SO_4^{2-}$), which on the modern Earth is an enormously important internal $CH_4$ sink within Earth's oceans (Egger et al., 2018). Their simulations did not explore atmospheric chemistry. Similarly, Daines and Lenton (2016) also innovated over traditional box modelling approaches by applying an ocean general circulation model (GCM) to examine the role of aerobic methanotrophy in modulating ocean-atmosphere fluxes of $CH_4$ during Archean time (prior to ~2.5 billion years ago, Ga). However, this analysis likewise did not include AOM, and the GCM results were not coupled to atmospheric chemistry. In contrast, Olson et al. (2016) included AOM in a 3-D ocean biogeochemistry model coupled to an atmospheric chemistry routine and found that AOM represents a critical internal $CH_4$ sink in the oceans even at relatively low dissolved $SO_4^{2-}$ levels. Though this represented an important further step forward in understanding marine $CH_4$ cycling on the early Earth, Olson et al. (2016) employed a simplified parameterization of aerobic $CH_4$ consumption, neglected the thermodynamics of $CH_4$-consuming metabolisms under energy-limited conditions, and employed a parameterization of atmospheric $O_2$-$O_3$-$CH_4$ photochemistry that is most readily applicable to only a subset of the atmospheric $pO_2$ values characteristic of Earth history (Daines and Lenton, 2016;Olson et al., 2016). While all of these studies provided new modelling innovations and advances in understanding, important facets of global $CH_4$ cycling, particularly as relevant to the evolution of early Earth, were lacking.

Here, we present a new framework for modeling the ocean-atmosphere biogeochemical $CH_4$ cycle in the 'muffin' release of the cGENIE Earth system model. Our goal is to make further progress in the development of a flexible intermediate-complexity model suitable for simulating the global biogeochemical $CH_4$ cycle on ocean-bearing planets, with an initial focus on periods of Earth history (or other habitable ocean-bearing planets) that lack a robust terrestrial biosphere. We also aim to provide a numerical modeling foundation from which to further develop a more complete $CH_4$ cycle within the cGENIE framework, including, for example, dynamic $CH_4$ hydrate cycling and the production/consumption of $CH_4$ by terrestrial ecosystems.

The outline of the paper is as follows. In Section 2 we briefly describe the GENIE/cGENIE Earth system model, with a particular eye toward the features that are most relevant for the biological carbon pump and the oceanic $CH_4$ cycle. In Section 3 we describe the major microbial metabolisms involved in the oceanic $CH_4$ cycle and compare our parameterizations to data from modern marine environments. In Section 4 we describe two alternative parameterizations of atmospheric $O_2$-$O_3$-$CH_4$ photochemistry incorporated into the model and compare these to modern/recent observations. In Section 5 we present results from a series of idealized simulations meant to illustrate the flexibility of the model and some potential applications. The availability of the model code, plus configuration files for all experiments described in the paper, is provided in Section 7, following a brief summary in Section 6.

## 2. The GENIE/cGENIE Earth system model
### 2.1. Ocean physics and climate model – C-GOLDSTEIN

The ocean physics and climate model in cGENIE is comprised of a reduced physics (frictional geostrophic) 3-D ocean circulation model coupled to both a 2-D energy-moisture balance model (EMBM) and a dynamic-thermodynamic sea-ice model (Edwards and Marsh, 2005;Marsh et al., 2011). The ocean model transports heat, salinity, and biogeochemical tracers using a scheme of parameterized isoneutral diffusion and eddy-induced advection (Griffies, 1998;Edwards and Marsh, 2005;Marsh et al., 2011), exchanges heat and moisture with the atmosphere, sea ice, and land, and is forced at the ocean surface by the input of zonal and meridional wind stress according to a specified wind field. The 2-D atmospheric energy-moisture-balance model (EMBM) considers the heat and moisture balance for the atmospheric boundary layer using air temperature and specific humidity as prognostic tracers. Heat and moisture are mixed horizontally throughout the atmosphere, and exchange heat and moisture with the ocean and land surfaces with precipitation occurring above a given relative humidity threshold. The sea-ice model tracks the horizontal transport of sea ice, and the exchange of heat and freshwater with the ocean and atmosphere using ice thickness, areal fraction, and concentration as prognostic variables. Full descriptions of the model and coupling procedure can be found in Edwards and Marsh (2005) and, more recently, in Marsh et al. (2011). As implemented here, the ocean model is configured as a 36 x 36 equal-area grid (uniform in longitude and uniform in the sine of latitude) with 16 logarithmically spaced depth levels and seasonal surface forcing from the EMBM.

**2.2. Ocean biological pump – BIOGEM**

The biogeochemical model component — 'BIOGEM' — regulates air-sea gas exchange as well as the transformation and partitioning of biogeochemical tracers within the ocean, as described in Ridgwell et al. (2007). By default, the biological pump is driven by parameterized uptake of nutrients in the surface ocean, with this flux converted stoichiometrically to biomass that is then partitioned into either dissolved or particulate organic matter for downstream transport, sinking, and remineralization. Dissolved organic matter is transported by the ocean model and decays with a specified time constant, while particulate organic matter is immediately exported out of the surface ocean and partitioned into two fractions of differing lability. In the ocean interior, particulate organic matter is remineralized instantaneously throughout the water column following an exponential decay function with a specified remineralization length scale.

In the simulations discussed below, photosynthetic nutrient uptake in surface ocean grid cells is controlled by a single limiting nutrient, dissolved phosphate ($PO_4$):

$$\frac{\partial PO_4}{\partial t} = -\Gamma + \lambda DOP, \tag{1}$$

$$\frac{\partial \text{DOP}}{\partial t} = \upsilon \Gamma - \lambda \text{DOP}, \tag{2}$$

where DOP represents dissolved organic phosphorus, $\upsilon$ represents the proportion of photosynthetic production that is initially partitioned into a dissolved organic phase, $\lambda$ represents a decay constant (time$^{-1}$) for dissolved organic matter, and $\Gamma$ represents photosynthetic nutrient uptake following Doney et al. (2006):

$$\Gamma = F_I \cdot F_N \cdot F_T \cdot (1 - f_{ice}) \cdot \frac{[\text{PO}_4]}{\tau_{bio}}. \tag{3}$$

Rates of photosynthesis are regulated by terms describing the impact of available light ($F_I$), nutrient abundance ($F_N$), temperature ($F_T$), and fractional sea ice coverage in each grid cell ($f_{ice}$). Rates of photosynthetic nutrient uptake are further scaled to ambient dissolved PO$_4$ ([PO$_4$]) according to an optimal uptake timescale ($\tau_{bio}$).

Note that this parameterization differs from that in Ridgwell et al. (2007). Specifically, the impacts of light and nutrient availability are both described via Michaelis-Menten terms:

$$F_I = \frac{I}{I + \kappa_I}, \tag{4}$$

$$F_N = \frac{[\text{PO}_4^{3-}]}{\kappa_P + [\text{PO}_4^{3-}]}, \tag{5}$$

where shortwave irradiance $I$ is averaged over the entire mixed layer, and is assumed to decay exponentially from the sea surface with a length scale of 20 m. It is assumed that nutrient uptake and photosynthetic production only occur in surface grid cells of cGENIE (e.g., the upper 80 m), which is similar to the 'compensation depth' $z_c$ in Doney et al. (2006) of 75 m. The terms $\kappa_I$ and $\kappa_P$ represent half-saturation constants for light and dissolved phosphate, respectively. In addition, the effect of temperature on nutrient uptake is parameterized according to:

$$F_T = k_{T0} \cdot \exp\left[\frac{T}{k_{eT}}\right], \tag{6}$$

where $k_{T0}$ and $k_{eT}$ denote pre-exponential and exponential scaling constants and $T$ represents absolute *in-situ* temperature. The scaling constants are chosen to give approximately a factor of two change in rate with a temperature change of 10°C (e.g., a Q$_{10}$ response of ~2.0). Lastly, the final term in Eq. (3), not present in the default parameterization of Ridgwell et al. (2007), allows for biological productivity to scale more directly with available PO4 when dissolved PO4 concentrations are elevated relative to those of the modern oceans.

Particulate organic matter (POM) is immediately exported out of the surface ocean without lateral advection, and is instantaneously remineralized throughout the water column according to an exponential function of depth:

$$F_z^{POM} = F_{z=z_h}^{POM} \cdot \left( \sum_i r_i^{POM} \cdot \exp\left( \frac{z_h - z}{l_i^{POM}} \right) \right), \tag{7}$$

where $F_z^{POM}$ is the particulate organic matter flux at a given depth (and $z_h$ is the base of the photic zone), $z$ is depth, $r_i^{POM}$ and $l_i^{POM}$ refer to the relative partitioning into each organic matter lability fraction $i$ and the $e$-folding depth of that fraction, respectively. The simulations presented here employ two organic matter fractions, a 'labile' fraction (94.5%) with an $e$-folding depth of ~590 m and an effectively inaccessible fraction (5.5%) with an $e$-folding depth of $10^6$ m (**Table 1**).

We employ a revised scheme for organic matter remineralization in the ocean interior, following that commonly used in models of organic matter remineralization within marine and lacustrine sediments (Rabouille and Gaillard, 1991; Van Cappellen et al., 1993; Boudreau, 1996a, b). Respiratory electron acceptors ($O_2$, $NO_3^-$, and $SO_4^{2-}$) are consumed according to decreasing free energy yield (Froelich et al., 1979), with consumption rates ($R_i$) scaled to both electron acceptor abundance and the inhibitory impact of electron acceptors with higher intrinsic free energy yield:

$$R_{O_2} = \frac{[O_2]}{\kappa_{O_2} + [O_2]}, \tag{8}$$

$$R_{NO_3} = \frac{[NO_3]}{\kappa_{NO_3} + [NO_3]} \cdot \frac{\kappa_{O_2}^i}{\kappa_{O_2}^i + [O_2]}, \tag{9}$$

$$R_{SO_4} = \frac{[SO_4]}{\kappa_{SO_4} + [SO_4]} \cdot \frac{\kappa_{O_2}^i}{\kappa_{O_2}^i + [O_2]} \cdot \frac{\kappa_{NO_3}^i}{\kappa_{NO_3}^i + [NO_3]}, \tag{10}$$

with the exception that in the biogeochemical configuration used here we do not consider nitrate ($NO_3^-$). The total consumption of settling POM within each ocean layer is governed by the predetermined remineralization profiles (Equation 7). The $R_i$ terms denote the relative fraction of this organic matter consumption that is performed by each respiratory process. We specify a closed system with no net organic matter burial in marine sediments (see below) and hence the POM flux to the sediment surface is assumed to be completely degraded, with the same partitioning amongst electron acceptors carried out according to local bottom water chemistry. For DOM, the assumed lifetime ($\lambda$) determines the total fraction of DOM degraded (and Equations 8-10 again determine how the consumption of electron acceptors is partitioned). The $\kappa_i$ terms represent half-saturation constants for each metabolism, $\kappa_i^i$ terms give inhibition constants acting on less energetic downstream respiratory processes, and brackets denote concentration. Default parameter values used here are shown in **Table 1**.

## 3. Oceanic methane cycling
### 3.1. Microbial methanogenesis
Methanogenesis represents the terminal step in our remineralization scheme, and follows the overall stoichiometry:

$$2CH_2O \rightarrow CH_4 + CO_2 .$$

This can be taken to implicitly include fermentation of organic matter to acetate followed by acetoclastic methanogenesis:

$$2CH_2O \rightarrow CH_3COOH ,$$
$$CH_3COOH \rightarrow CH_4 + CO_2 ,$$

or the fermentation of organic matter to acetate followed by anaerobic acetate oxidation and hydrogenotrophic methanogenesis:

$$2CH_2O \rightarrow CH_3COOH ,$$
$$CH_3COOH + 2H_2O \rightarrow 4H_2 + 2CO_2 ,$$
$$4H_2 + CO_2 \rightarrow CH_4 + 2H_2O ,$$

both of which have the same overall net stoichiometry provided that $H_2$ is assumed to be quantitatively converted to $CH_4$ by hydrogenotrophic methanogens. We thus ignore the scenario in which some fraction of $H_2$ is converted directly to biomass by hydrogenotrophic methanogens acting as primary producers (Ozaki et al., 2018).

Because we specify a closed system with no net organic matter burial in marine sediments, all organic matter not remineralized by more energetic respiratory metabolisms is converted into CH$_4$ (e.g., $R_{CH4} = 1 - R_{O2} - R_{NO3} - R_{SO4}$):

$$R_{CH_4} = \frac{\kappa^i_{O_2}}{\kappa^i_{O_2}+[O_2]} \cdot \frac{\kappa^i_{NO_3}}{\kappa^i_{NO_3}+[NO_3]} \cdot \frac{\kappa^i_{SO_4}}{\kappa^i_{SO_4}+[SO_4]}, \qquad (11)$$

where $\kappa_i$ and $\kappa_i^i$ terms are as described above (**Table 1**). We disable nitrate (NO$_3$) as a tracer in the simulations presented here, such that anaerobic remineralization of organic matter is partitioned entirely between sulfate reduction and methanogenesis (**Fig. 1**). Using our default parameter values (**Table 1**), aerobic respiration dominates organic matter remineralization at [O$_2$] values significantly above 1 µmol kg$^{-1}$ (**Fig. 1a**) while anaerobic remineralization is dominated by methanogenesis at [SO$_4^{2-}$] values significantly below 1 mmol kg$^{-1}$ (**Fig. 1b**). An important outcome of the revised 'inhibition' scheme is that metabolic pathways with differing intrinsic free energy yields can coexist, which more accurately reflects field observations from a range of natural settings (Curtis, 2003;Bethke et al., 2008;Kuivila et al., 1989;Jakobsen and Postma, 1999). In particular, it allows us to roughly capture the impact of oxidant gradients within sinking marine aggregates (Bianchi et al., 2018), which can facilitate non-trivial anaerobic carbon remineralization within sinking particles even in the presence of relatively high [O$_2$] in the ocean water column (**Fig. 1c**).

While the model tracks the carbon isotope composition of oceanic and atmospheric CH$_4$ ($\delta^{13}$C, reported in per mil notation relative to the Pee Dee Belemnite, PDB), the only significant isotope effect we include here is that attendant to acetoclastic methanogenesis. We specify a constant isotope fractionation between organic carbon and CH$_4$ during methanogenesis of -35‰ by default (**Table 2**), which will tend to produce microbial CH$_4$ with a $\delta^{13}$C composition of roughly -60‰ when combined with the default isotope fractionation associated with photosynthetic carbon fixation in the surface ocean (e.g., Kirtland Turner and Ridgwell, 2016). The model does not currently include any potential isotope effects associated with aerobic/anaerobic methanotrophy, air-sea gas exchange of CH$_4$, or photochemical breakdown of CH$_4$ in the atmosphere. It does, however, include a comprehensive $^{13}$C scheme associated with ocean-atmosphere cycling of CO$_2$ (Kirtland Turner and Ridgwell, 2016;Ridgwell, 2001).

## 3.2. Aerobic methanotrophy

Microbial aerobic methanotrophy proceeds according to:

$$CH_4 + 2O_2 \rightarrow CO_2 + 2H_2O$$

This reaction is highly favorable energetically, with a free energy yield under standard conditions of ~850 kJ per mole of methane consumed (**Table 2**). We represent rates of aerobic methanotrophy ($R_{AER}$) with a mixed kinetic-thermodynamic formulation (Jin & Bethke, 2005; 2007; Regnier et al., 2011), in which $CH_4$ oxidation kinetics are controlled by substrate availability, thermodynamic energy yield, and temperature:

$$R_{AER} = k_{AER} \cdot F_k^{AER} \cdot F_t^{AER} \cdot F_T . \qquad (12)$$

A rate constant for aerobic methanotrophy ($y^{-1}$) is defined as $k_{AER}$, while $F_i$ terms denote kinetic ($k$) and thermodynamic ($t$) factors as defined below and a temperature ($T$) factor as given in Eq. (6) above.

The kinetic factor ($F_k$) for aerobic methanotrophy is controlled by substrate availability according to:

$$F_k^{AER} = [CH_4] \cdot \frac{[O_2]}{\kappa_O^{AER} + [O_2]} , \qquad (13)$$

where brackets denote concentration and the $\kappa$ term denotes a half-saturation constant with respect to $O_2$. We employ a hybrid parameterization in which kinetics are first-order with respect to $CH_4$ but also scaled by a Michaelis-Menten-type term for $O_2$. This formulation is based on the rationale that half-saturation constants for $CH_4$ are typically similar to (or greater than) the dissolved $CH_4$ levels attained in anoxic water column environments (Regnier et al., 2011) but is also meant to allow for rapid $CH_4$ consumption under 'bloom' conditions with an appropriately scaled rate constant (see below).

The effect of thermodynamic energy yield on aerobic methanotrophy is given by:

$$F_t^{AER} = 1 - \exp\left[\frac{\Delta G_{r,AER} + \Delta G_{BQ,AER}}{\chi RT}\right], \qquad (14)$$

where $\Delta G_r$ denotes the Gibbs free energy of reaction under *in-situ* conditions, $\Delta G_{BQ}$ represents the minimum energy required to sustain ATP synthesis (Hoehler et al., 2001;Hoehler, 2004;Jin and Bethke, 2007), χ is the stoichiometric number of the reaction (e.g., the number of times the rate-determining step occurs in the overall process), and $R$ and $T$ represent the gas constant and absolute *in-situ* temperature, respectively. The available free energy is estimated according to:

$$\Delta G_{r,AER} = \Delta G^0_{r,AER} + RT \cdot \ln \frac{\gamma_{CO_2}[CO_2]}{\gamma_{O_2}[O_2] \cdot \gamma_{CH_4}[CH_4]} , \qquad (15)$$

where, in addition to the terms defined above, $\Delta G_r^0$ represents the Gibbs free energy of the reaction under standard conditions, and $\gamma_i$ values represent activity coefficients. Note that we assume an $H_2O$ activity of unity.

### 3.3. Anaerobic oxidation of methane (AOM)

The oxidation of methane can also be coupled to electron acceptors other than $O_2$, including nitrate ($NO_3^-$), sulfate ($SO_4^{2-}$), and oxide phases of iron (Fe) and manganese (Mn) (Reeburgh, 1976; Martens and Berner, 1977; Hoehler et al., 1994; Hinrichs et al., 1999; Orphan et al., 2001; Sivan et al., 2011; Haroon et al., 2013; Egger et al., 2015). Because it is by far the most abundant of these oxidants on the modern Earth, and has likely been the most abundant throughout Earth's history, we focus on anaerobic oxidation of methane (AOM) at the expense of $SO_4^{2-}$:

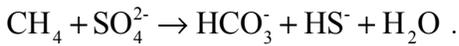
$$CH_4 + SO_4^{2-} \rightarrow HCO_3^- + HS^- + H_2O .$$

This process is currently thought to be performed most often through a syntrophic association between Archaea and sulfate reducing bacteria (Boetius et al., 2000), though the mechanics controlling the exchange of reducing equivalents within the syntrophy remain to be fully elucidated (Milucka et al., 2012; McGlynn et al., 2015). In any case, consumption of $CH_4$ at the sulfate-methane transition zone (SMTZ) represents an extremely large sink flux of $CH_4$ in modern marine sediments (Regnier et al., 2011; Egger et al., 2018).

Anaerobic methanotrophy is much less energetically favorable under standard conditions, with a free energy yield of ~30 kJ per mole of $CH_4$ (**Table 2**). As a result, the influence of thermodynamics on rates of AOM is potentially much stronger than it will tend to be in the case of aerobic methanotrophy. As above, rates of AOM are controlled by the combined influence of substrate availability, thermodynamic drive, and temperature:

$$R_{AOM} = k_{AOM} \cdot F_k^{AOM} \cdot F_t^{AOM} \cdot F_T \qquad (16)$$

where $k_{AOM}$ is a rate constant for anaerobic methane oxidation (y$^{-1}$), while $F_i$ terms denote kinetic ($k$) and thermodynamic ($t$) factors as defined below and a temperature ($T$) factor as given in Eq. (6) above.

The kinetics of anaerobic methane oxidation are specified according to:

$$F_k^{AOM} = [CH_4] \cdot \frac{[SO_4^{2-}]}{\kappa_S^{AOM} + [SO_4^{2-}]} \qquad (17)$$

where brackets denote concentration and the $\kappa$ term denotes a half-saturation constant with respect to $SO_4^{2-}$. We employ a hybrid parameterization in which kinetics are first-order with respect to $CH_4$ but are also scaled by a Michaelis-Menten-type term for $SO_4^{2-}$ for reasons discussed above.

The effect of thermodynamic energy yield on anaerobic methane oxidation is specified as follows:

$$F_t^{AOM} = 1 - \exp\left[\frac{\Delta G_{r,AOM} + \Delta G_{BQ,AOM}}{\chi RT}\right] \qquad (18)$$

As above, $\Delta G_r$ denotes the Gibbs free energy of reaction under *in-situ* conditions, $\Delta G_{BQ}$ is the minimum energy required to sustain ATP synthesis (the 'biological quantum'), $\chi$ is the stoichiometric number of the reaction, and $R$ and $T$ represent the gas constant and absolute *in-situ* temperature, respectively. The available free energy for AOM under *in-situ* conditions is estimated according to:

$$\Delta G_{r,AOM} = \Delta G_{r,AOM}^0 + RT \cdot \ln \frac{\gamma_{HCO_3^-}[HCO_3^-] \cdot \gamma_{HS^-}[HS^-]}{\gamma_{SO_4^{2-}}[SO_4^{2-}] \cdot \gamma_{CH_4}[CH_4]}, \qquad (19)$$

where $\Delta G_r^0$ again represents the Gibbs free energy of the net AOM reaction given above under standard conditions, and $\gamma_i$ values represent activity coefficients. Again, we assume an H$_2$O activity of unity.

### 3.4. Default parameters for aerobic and anaerobic methanotrophy

We choose default rate constants according to a dataset of compiled rates of aerobic and anaerobic methanotrophy in oxygenated and anoxic marine water column environments (see Supplementary Data), after correction to *in-situ* temperature (**Fig. 2a, b**). Our default values for both rate constants are on the low end of the observational dataset, but are very roughly tuned to yield steady-state diffusive CH$_4$ fluxes from the ocean that are consistent with recent observational constraints (**Fig. 2c**). It is important to note, however, that these values are not extensively tuned and could be adjusted depending on the application. For example, transient CH$_4$ release experiments could employ rate constants that are scaled upward to reflect transient ('bloom') elevations in microbial community CH$_4$ consumption as observed in field studies (Kessler et al., 2011;Crespo-Medina et al., 2014). Default values for other kinetic parameters (**Table 2**) are chosen to be broadly consistent with field measurements and pure/mixed culture experiments with aerobic methanotrophs (Bender and Conrad, 1992, 1993;Hanson and Hanson, 1996;Dunfield and Conrad, 2000;van Bodegom et al., 2001), and to remain roughly consistent with previous work for comparative purposes (e.g., Olson et al., 2016), though the parameters have not been formally tuned and we explore model sensitivity below.

Thermodynamic energy yields of each reaction under standard conditions are calculated based on the standard molal thermodynamic properties given in Regnier et al. (2011). Stoichiometric numbers are assumed identical for both metabolisms, with default values of 1.0 (Jin and Bethke, 2005;Dale et al., 2006). We assume a default biological quantum ($\Delta G_{BQ}$) of 15 kJ mol$^{-1}$ for both aerobic and anaerobic methanotrophy, though these can be expected to vary somewhat as a function of metabolism and environmental conditions (Schink, 1997;Hoehler, 2004;Dale et al., 2008). These can be varied independently for aerobic and anaerobic methanotrophy in the model, and we explore model sensitivity to this parameter below. Lastly, for simplicity and to minimize computational expense we assume constant activity coefficients for each species throughout all ocean grid cells (**Table 2**). For some applications it may ultimately be important to add a scheme for estimating activity coefficients according to ambient salinity and ion chemistry, for example estimating methane fluxes in planetary scenarios with very different major ion chemistry or much higher/lower salinity than those characteristic of Earth's modern oceans.

## 4. Atmospheric methane cycling
### 4.1. Air-sea gas exchange
Ocean-atmosphere fluxes of CH$_4$ ($J_{gas}$) are governed by temperature- and salinity-dependent solubility and surface wind speed above a given grid cell:

$$J_{gas} = A \cdot k_{gas} \cdot \left( [CH_4]_{sat} - [CH_4]_{cell} \right), \tag{20}$$

where $A$ denotes the area available for gas-exchange (e.g., the area of ice-free surface ocean), $[CH_4]_{cell}$ denotes the ambient dissolved $CH_4$ concentration in a given surface ocean grid cell, $[CH_4]_{sat}$ represents the dissolved $CH_4$ concentration at saturation with a given atmospheric $pCH_4$, temperature, and salinity, and $k_{gas}$ represents a gas transfer velocity. Solubility is based on a Bunsen solubility coefficient ($\beta$) corrected for ambient temperature ($T$) and salinity ($S$) according to:

$$\ln \beta = a_1 + a_2 (100/T) + a_3 \ln(T/300) + S \left[ b_1 + b_2 (T/100) + b_3 (T/100)^2 \right], \tag{21}$$

[Note that the Henry's law constant $K_0$ is related to the Bunsen solubility coefficient by $K_0 = \beta/\rho V^+$, where $\rho$ is density and $V^+$ is the molar volume of the gas at STP.] Gas transfer velocity ($k_{gas}$) is calculated based on the surface windspeed ($u$) and a Schmidt number (Sc) corrected for temperature assuming a constant salinity of 35‰:

$$k_{gas} = k \cdot u^2 \cdot \left[ Sc/660 \right]^{-0.5}, \tag{22}$$

where $k$ is a dimensionless gas transfer coefficient, $u$ is surface wind speed, and Sc is the temperature-corrected Schmidt number according to:

$$Sc = c_1 - c_2 T + c_3 T^2 - c_4 T^3. \tag{23}$$

All default constants and coefficients for the gas exchange scheme are given in **Table 3**. Overall, the scheme for air-sea gas exchange of $CH_4$ follows by default that for other gases accounted for in BIOGEM, such as $O_2$ and $CO_2$, as described in (Ridgwell et al., 2007)

### 4.2. Parameterized $O_2$-$O_3$-$CH_4$ photochemistry

Once degassed to the atmosphere, $CH_4$ becomes involved in a complex series of photochemical reactions initiated by hydroxyl radical (OH) attack on $CH_4$ (Kasting et al., 1983;Prather, 1996;Pavlov et al., 2000;Schmidt and Shindell, 2003). Following Claire et al. (2006) and Goldblatt et al. (2006), we parameterize $O_2$-$O_3$-$CH_4$ photochemistry according to a bimolecular 'rate law':

$$J_{CH_4} = k_{eff} \cdot M_{O_2} \cdot M_{CH_4}, \tag{24}$$

where $M_i$ terms represent the atmospheric inventories of $O_2$ and $CH_4$, respectively, and $k_{eff}$ denotes an effective rate constant (Tmol$^{-1}$ y$^{-1}$) that is itself a complicated function of atmospheric $O_2$, $CH_4$, and $CO_2$ (Claire et al., 2006). At each timestep, the distribution of chemical species (e.g., other than temperature and humidity) in the atmosphere is homogenized (Ridgwell et al., 2007) and $k_{eff}$ is estimated based on the resulting instantaneous mean partial pressures of $O_2$ and $CH_4$ according to a bivariate fit to a large suite of 1-D atmospheric photochemical models. These photochemical model results (Claire, *personal communication*) are derived following Claire et al. (2006). Briefly, values for $k_{eff}$ are computed by a 1-D model of atmospheric photochemistry assuming a range of fixed surface mixing ratios of $O_2$ and $CH_4$ and a constant atmospheric $CO_2$ of $10^{-2}$ bar. We then fit a fifth-order polynomial surface to these $k_{eff}$ values as a function of atmospheric $pO_2$ and $pCH_4$ (**Fig. 3**).

Our default parameterization of $O_2$-$O_3$-$CH_4$ chemistry (C06) is fit over a $pO_2$ range of $10^{-14}$ to $10^{-1}$ bar, a $pCH_4$ range of $10^{-6}$ to $2 \times 10^{-3}$ bar, and a constant high background $pCO_2$ of $10^{-2}$ bar (Claire et al., 2006). We thus truncate the atmospheric lifetime of $CH_4$ at a lower bound of 7.6 years in our default parametrization, and provide an alternative parameterization of photochemical $CH_4$ destruction at roughly modern $pO_2$ and $pCO_2$ (SS03) derived from the results of Schmidt and Shindell (2003) for use in more geologically recent, high-$O_2$ atmospheres (Reinhard et al., 2017) (**Fig. 4a**). Although this parameterized photochemistry scheme should represent an improvement in accuracy relative to that implemented in Olson et al. (2016) (see Daines and Lenton, 2016), it is important to point out that a range of factors that might be expected to impact the photochemical destruction rates of $CH_4$ in the atmosphere, including atmospheric $pCO_2$, the atmospheric profile of $H_2O$, and spectral energy distribution (SED), have not yet been rigorously assessed. Ongoing model developments in ATCHEM are aimed at implementing a more flexible and inclusive photochemical parameterization that will allow for robust use across a wider range of atmospheric compositions and photochemical environments.

As a basic test of our photochemical parameterization, we impose a terrestrial (wetland) flux of $CH_4$ to the atmosphere (balanced by stoichiometric consumption of $CO_2$ and release of $O_2$), and allow the oceanic and atmospheric $CH_4$ cycle to spin up for 20 kyr. We then compare steady-state atmospheric $pCH_4$ as a function of terrestrial $CH_4$ flux to estimates for the last glacial, preindustrial, and modern periods. Our default parameterization is relatively simple and spans a very wide range in atmospheric $O_2$ and $CH_4$ inventories. Nevertheless, both the default scheme and the alternative parameterization for recent geologic history (and analogous planetary environments) with high-$pO_2$/low-$pCO_2$ atmospheres accurately reproduce atmospheric $pCH_4$ values given estimated glacial, preindustrial, and modern terrestrial $CH_4$ fluxes (**Fig. 4C**), and both display the predicted saturation of $CH_4$ sinks at elevated atmospheric $CH_4$ observed in more complex photochemical models. We note, however, the alternative parameterization tends to yield slightly higher atmospheric $pCH_4$ at surface fluxes greater than ~50 Tmol y$^{-1}$ (**Fig. 4C**). (In the remainder of the manuscript, we employ the default (C06) parameterization for atmospheric $O_2$-$O_3$-$CH_4$ chemistry and do not discuss the simple high-$pO_2$/low-$pCO_2$ alternative further.)

## 5. Example applications of the new capabilities in the cGENIE model
### 5.1. High-$pO_2$ ('modern') steady state

We explore a roughly modern steady state with appropriate continental geography and simulated overturning circulation (as in Cao et al., 2009) and initialize the atmosphere with $pO_2$, $pCO_2$, and $pCH_4$ of [v/v] 20.95%, 278 ppm, and 700 ppb, respectively, and with globally uniform oceanic concentrations of $SO_4^{2-}$ (28 mmol kg$^{-1}$) and $CH_4$ (1 nmol kg$^{-1}$). We fix globally averaged solar insolation at the modern value (1368 W m$^{-2}$) with seasonally variable forcing as a function of latitude, and set radiative forcing for $CO_2$ and $CH_4$ equivalent to preindustrial values in order to isolate the effects of biogeochemistry on steady state tracer distributions. The model is then spun up for 20 kyr with atmospheric $pO_2$ and $pCO_2$ (and $\delta^{13}C$ of atmospheric $CO_2$) restored to preindustrial values at every timestep, and with an imposed wetland flux of $CH_4$ to the atmosphere of 20 Tmol yr$^{-1}$ that has a $\delta^{13}C$ value of -60‰. Atmospheric $pCH_4$ and all oceanic tracers are allowed to evolve freely.

Surface, benthic, and ocean interior distributions of dissolved oxygen ($O_2$), sulfate ($SO_4^{2-}$), and methane ($CH_4$) are shown in **Fig. 5** for our roughly modern simulation. Dissolved $O_2$ ($[O_2]$) approaches air saturation throughout the surface ocean, with a distribution that is largely uniform zonally and with concentrations that increase with latitude as a result of increased solubility at lower temperature near the poles (**Fig. 5a**). Benthic $[O_2]$ shows patterns similar to those expected for the modern Earth, with relatively high values in the well-ventilated deep North Atlantic, low values in the deep North Pacific and Indian oceans, and a gradient between roughly air saturation near regions of deep convection in the high-latitude Atlantic and much lower values in the tropical and northern Pacific (**Fig. 5d**). Distributions of $[O_2]$ in the ocean interior are similar to those of the modern Earth (**Fig. 5g**) with oxygen minimum zones (OMZs) at intermediate depths underlying highly productive surface waters, particularly in association with coastal upwelling at low latitudes.

Concentrations of dissolved $SO_4^{2-}$ ($[SO_4^{2-}]$) are largely invariant throughout the ocean, consistent with its expected conservative behavior in the modern ocean as one of the most abundant negative ions in seawater (**Fig. 5**). Slightly higher concentrations in both surface and benthic fields are seen in association with outflow from the Mediterranean, and are driven by evaporative concentration (**Fig. 5b**). Benthic $[SO_4^{2-}]$ distributions show some similarity to those of $[O_2]$ (**Fig. 5e**), though again the differences are very small relative to the overall prescribed initial tracer inventory of 28 mmol kg$^{-1}$ and disappear almost entirely when salinity-normalized (not shown). In the ocean interior, $[SO_4^{2-}]$ is largely spatially invariant with a value of approximately 28 mmol kg$^{-1}$ (**Fig. 5h**).

Dissolved $CH_4$ concentrations ($[CH_4]$) in the surface and shallow subsurface ocean are much more variable, but typically on the order of ~1-2 nmol kg$^{-1}$ with slightly elevated concentrations just below the surface, both of which are consistent with observations from the modern ocean (Reeburgh, 2007;Scranton and Brewer, 1978). The benthic $[CH_4]$ distribution shows locally elevated values up to ~300-400 nmol kg$^{-1}$ in shallow regions of the tropical and northern Pacific and the Indian oceans (**Fig. 5f**), which is also broadly consistent with observations from shallow marine environments with active benthic $CH_4$ cycling (Jayakumar et al., 2001). Within the ocean interior, dissolved $CH_4$ can accumulate in the water column in excess of ~100 nmol kg$^{-1}$ in association with relatively low-$O_2$ conditions at intermediate depths, with zonally averaged values as high as ~70 nmol kg$^{-1}$ but more typically in the range of ~20-40 nmol kg$^{-1}$ (**Fig. 5i**). These concentrations are comparable to those observed locally in low-$O_2$ regions of the modern ocean (Sansone et al., 2001;Chronopoulou et al., 2017;Thamdrup et al., 2019).

**Figure 6** shows the major metabolic fluxes within the ocean's microbial $CH_4$ cycle for our 'modern' configuration. Methanogenesis is focused in regions characterized by relatively low $[O_2]$ and is particularly vigorou s in the Eastern Tropical Pacific, the North Pacific, and the Indian Ocean (**Fig. 6a**). The highest zonally averaged rates of methanogenesis are observed in northern tropical and subtropical latitudes, and are focused at a depth of ~1 km (**Fig. 6d**). Rates of microbial $CH_4$ consumption are generally spatially coupled to rates of methanogenesis, both in a column-integrated sense (**Fig. 6b, c**) and in the zonal average (**Fig. 6e, f**). This is particularly true for AOM, rates of which are highest within the core of elevated methanogenesis rates observed in the northern subtropics. Zonally averaged AOM rates of ~10-15 nmol kg$^{-1}$ d$^{-1}$ compare well with field measurements of AOM within oceanic OMZs (Thamdrup et al., 2019). In general, the bulk of CH$_4$ produced via microbial methanogenesis is consumed via AOM, either near the seafloor or within the ocean interior.

**5.2. Low-$p$O$_2$ ('ancient') steady state**

Next, we explore a low-$p$O$_2$ steady state, similar to the Proterozoic Earth (Reinhard et al., 2017) but played out in a modern continental configuration and overturning circulation, by initializing the atmosphere with $p$O$_2$, $p$CO$_2$, and $p$CH$_4$ of [v/v] 2.1 x 10$^{-4}$ atm (equivalent to a value 10$^{-3}$ times the present atmospheric level, PAL), 278 ppm, and 500 ppm, respectively, and globally uniform oceanic concentrations of SO$_4^{2-}$ (280 µmol kg$^{-1}$) and CH$_4$ (50 µmol kg$^{-1}$). We again fix globally averaged solar insolation at the modern value (1368 W m$^{-2}$) with seasonally variable forcing as a function of latitude, and set radiative forcing for CO$_2$ and CH$_4$ equivalent to the modern preindustrial state in order to isolate the effects of biogeochemistry on steady state tracer distributions. The model is then spun up for 20 kyr with atmospheric $p$O$_2$ and $p$CO$_2$ (and δ$^{13}$C of atmospheric CO$_2$) restored to the initial values specified above at every timestep, with an imposed 'geologic' flux of CH$_4$ to the atmosphere of 3 Tmol yr$^{-1}$ at a δ$^{13}$C value of -60‰. Atmospheric $p$CH$_4$ and all oceanic tracers are allowed to evolve freely.

Surface, benthic, and ocean interior distributions of [O$_2$], [SO$_4^{2-}$], and [CH$_4$], are shown in **Fig. 7** for our low-$p$O$_2$ simulation. Dissolved O$_2$ concentrations are now extremely heterogeneous throughout the surface ocean, ranging over an order of magnitude from less than 1 µmol kg$^{-1}$ to over 10 µmol kg$^{-1}$, with concentrations that are regionally well in excess of air saturation at the prescribed $p$O$_2$ of 2.1 x 10$^{-4}$ atm (**Fig. 7a**). Previous studies have shown that these features are not unexpected at very low atmospheric $p$O$_2$ (Olson et al., 2013; Reinhard et al., 2016). We note, however, that the distribution and maximum [O$_2$] in our low-$p$O$_2$ simulation are both somewhat different from those presented in Olson et al. (2013) and Reinhard et al. (2016). We attribute this primarily to the different parameterizations of primary production in the surface ocean. In the biogeochemical configuration of cGENIE we adopt here, we allow rates of photosynthesis to scale more directly with available $PO_4^{3-}$ than is the case in these previous studies (Eq. 3), which allows for higher rates of oxygen production in regions of deep mixing and relatively intense organic matter recycling below the photic zone (**Fig. 7a**). In any case, as in previous examinations of surface [$O_2$] dynamics at low atmospheric $pO_2$ (Olson et al., 2013;Reinhard et al., 2016), our regional [$O_2$] patterns still generally track the localized balance between photosynthetic $O_2$ release and consumption through respiration and reaction with inorganic reductants, rather than temperature-dependent solubility patterns (**Fig. 5a**). Within the ocean interior, $O_2$ is consumed within the upper few hundred meters and is completely absent in benthic settings (**Fig. 7d, g**).

In our low-$pO_2$ simulations we initialize the ocean with a globally uniform [$SO_4^{2-}$] of 280 μmol kg$^{-1}$, under the premise that marine $SO_4^{2-}$ inventory should scale positively with atmospheric $pO_2$. With this much lower initial $SO_4^{2-}$ inventory (i.e., $10^2$ times less than the modern ocean), steady state [$SO_4^{2-}$] distributions are significantly more heterogeneous than in the modern, high-$pO_2$ case (**Fig. 7**). Ocean [$SO_4^{2-}$] is approximately homogeneous spatially in surface waters, even with a significantly reduced seawater inventory (**Fig. 7a**), but is strongly variable within the ocean interior (**Fig. 7e, h**). Indeed, in our low-$pO_2$ simulations $SO_4^{2-}$ serves as the principal oxidant for organic matter remineralization in the ocean interior, with the result that its distribution effectively mirrors that of [$O_2$] in the modern case in both spatial texture and overall magnitude (compare **Fig. 7e, h** with **Fig. 5d, g**). Dissolved $SO_4^{2-}$ in this simulation never drops to zero, a consequence of our initial 280 μmol kg$^{-1}$ concentration of $SO_4^{2-}$ representing the oxidative potential of 560 μmol kg$^{-1}$ of $O_2$, some 3 times higher than the mean [$O_2$] value in the modern ocean interior (~170 μmol kg$^{-1}$).

Dissolved $CH_4$ concentrations in the surface and shallow subsurface ocean are variable but much higher than in our modern simulations, typically on the order of ~1-2 μmol kg$^{-1}$ (**Fig. 7c**). The benthic [$CH_4$] distribution shows concentrations up to ~8 μmol kg$^{-1}$, with concentrations in excess of 1 μmol kg$^{-1}$ pervasively distributed across the seafloor. In general, the benthic [$CH_4$] distribution inversely mirrors that of [$SO_4^{2-}$] (**Fig. 7f**), which results from the fact that in the low-$pO_2$ case $SO_4^{2-}$ again serves as the principal oxidant of methane. Concentrations of $CH_4$ in the ocean interior can approach ~10 μmol kg$^{-1}$, but in the zonal average are typically less than 5 μmol kg$^{-1}$ (**Fig. 7i**). Overall, the oceanic $CH_4$ inventory increases dramatically in the low-$pO_2$ case relative to the modern simulation, from ~4.5 Tmol $CH_4$ to ~1900 Tmol $CH_4$.

**Figure 8** shows the major metabolic fluxes within the ocean's microbial $CH_4$ cycle for our 'ancient' configuration. Column-integrated rates of microbial methanogenesis are greater than in the high-$pO_2$ case by up to a factor of ~$10^2$ (**Fig. 8a**), with methanogenesis also showing a much broader areal distribution. Within the ocean interior, rates of methanogenesis are most elevated in the upper ~1 km (**Fig. 8d**) as a consequence of elevated rates of organic carbon remineralization combined with a virtual absence of dissolved $O_2$ beneath the upper ~200 m. Rates of aerobic methanotrophy, which is effectively absent in the ocean interior (**Fig. 8e**), are elevated relative to those observed the high-$pO_2$ simulation by less than an order of magnitude and are concentrated in the tropical surface ocean near the equatorial divergence (**Fig. 8b**). In contrast, AOM is strongly coupled spatially to microbial methanogenesis, with rates that are often well over ~$10^2$ times higher than those observed in the high-$pO_2$ case (**Fig. 8c, f**). Once again, AOM dominates the consumption of $CH_4$ produced in the ocean interior and acts as an extremely effective throttle on $CH_4$ fluxes to the atmosphere. Despite a significant increase in overall oceanic $CH_4$ burden relative to our high-$pO_2$ simulation (see above and **Fig. 7i**), atmospheric $pCH_4$ increases only modestly from ~0.8 ppm to 6 ppm [v/v], equivalent to an additional radiative forcing of only ~2 W m$^{-2}$, due to efficient microbial consumption in the upper ocean.

### 5.3. Atmospheric carbon injection

To illustrate the capabilities of the model in exploring the time-dependent (perturbation) behavior of the $CH_4$ cycle, we perform a simple carbon injection experiment in which 3,000 PgC are injected directly into the atmosphere either as $CH_4$ or as $CO_2$, starting from our modern steady state. The injection is spread over 1,000 years, with an instantaneous initiation and termination of carbon input to the atmosphere. The magnitude and duration of this carbon injection, corresponding to 3 PgC y$^{-1}$, is meant to roughly mimic the upper end of estimates for the Paleocene-Eocene Thermal Maximum, a transient global warming event at ~56 Ma hypothesized to have been driven by emissions of $CO_2$ and/or $CH_4$ (Kirtland Turner, 2018). This flux is much lower than the current anthropogenic carbon input of ~10 PgC y$^{-1}$ (Ciais et al., 2013). For simplicity, and because we focus on only the first 3,000 years following carbon injection, we treat the ocean-atmosphere system as closed, with the result that all injected carbon ultimately accumulates within the ocean and atmosphere rather than being removed through carbonate compensation and silicate weathering.

Following a carbon release to the atmosphere in the form of $CH_4$, there is an immediate and significant increase in atmospheric $pCH_4$ to values greater than 10 ppmv, followed by a gradual increase to a maximum of ~12 ppmv throughout the duration of the $CH_4$ input (**Fig. 9a**). Much of this methane is exchanged with the surface ocean and consumed by aerobic methanotrophy, while some is photochemically oxidized directly in the atmosphere, both of which lead to a significant but delayed increase in atmospheric $pCO_2$ (**Fig. 9b**). This increase in atmospheric $pCH_4$ and $pCO_2$ leads to an increase in global average surface air temperature (SAT) of ~7°C (**Fig. 9d**), an increase in mean ocean temperature (MOT) of ~2°C (**Fig. 9e**), along with significant acidification of the surface ocean (**Fig. 9c**).

The increase in atmospheric $pCO_2$ and drop in ocean pH are nearly identical if we instead inject the carbon as $CO_2$ rather than $CH_4$. (**Fig. 9b, c**). However, when carbon is injected as $CH_4$, there is an additional transient increase in global surface air temperature of ~2°C and roughly 0.5°C of additional whole ocean warming for the same carbon input and duration (**Fig. 9f**). This results from the fact that mole-for-mole, $CH_4$ is a much more powerful greenhouse gas than is $CO_2$, and oxidation of $CH_4$ to $CO_2$ is not instantaneous during the carbon release interval. Combined, these factors result in a disequilibrium situation in which a proportion of carbon released to the atmosphere remains in the form of $CH_4$ rather than $CO_2$, providing an enhancement of warming, especially during the duration of carbon input. This warming enhancement should be considered in past events during which $CH_4$ release is suspected as a key driver of warming. For instance, additional warming due to $CH_4$ forcing may help explain the apparent discrepancy between the amount of warming reconstructed by proxy records and proposed carbon forcing during the PETM (Zeebe et al., 2009)

## 5.4. Atmospheric $pCH_4$ on the early Earth

Using our low-$pO_2$ steady state as a benchmark case (**Section 5.2**), we briefly explore the sensitivity of atmospheric $pCH_4$ to a subset of model variables. All model ensembles are initially configured with globally homogeneous marine $SO_4^{2-}$ and $CH_4$ inventories and a background geologic $CH_4$ flux of 3 Tmol y$^{-1}$, and are spun up for 20 kyr with a fixed $pO_2$ and $pCO_2$. We report atmospheric $pCH_4$ from the final model year. Our purpose here is not to be exhaustive or to elucidate any particular period of Earth history, but to demonstrate some of the major factors controlling the atmospheric abundance of $CH_4$ on a low-oxygen Earth-like planet. We present results from individual sensitivity ensembles from our benchmark low-$pO_2$ case over the following parameter ranges: (1) atmospheric $pO_2$ between $10^{-4}$ to $10^{-1}$ times the present atmospheric level (PAL), equivalent to roughly 2 x $10^{-5}$ and 2 x $10^{-2}$ atm, respectively; (2) initial marine $SO_4^{2-}$ inventories corresponding to globally uniform seawater concentrations between 0 and 1,000 μmol kg$^{-1}$; and (3) biological energy quanta (BEQ) for anaerobic methane oxidation between 5 and 30 kJ mol$^{-1}$.

Results for our low-$pO_2$ sensitivity ensembles are shown in **Figure 10**. We find a similar sensitivity of atmospheric $pCH_4$ to atmospheric $pO_2$ to that observed by (Olson et al., 2016). In particular, atmospheric $CH_4$ abundance initially increases as atmospheric $pO_2$ drops below modern values to roughly 2-3% PAL, after which decreasing $pO_2$ causes $pCH_4$ to drop. This behavior is well-known from previous 1-D photochemical model analysis, and arises principally from increasing production of OH via water vapor photolysis as shielding of $H_2O$ by ozone ($O_3$) decreases at low atmospheric $pO_2$ (Pavlov et al., 2003;Claire et al., 2006;Goldblatt et al., 2006). However, peak atmospheric $pCH_4$ is significantly reduced in our models relative to those of Olson et al. (2016). For example, at an 'optimal' atmospheric $pO_2$ of ~2.5% PAL Olson et al. (2016) predict a steady state atmospheric $pCH_4$ of ~35 ppmv, while we predict a value of ~10 ppmv (**Fig. 10a**). This difference can be attributed to our updated $O_2$-$O_3$-$CH_4$ photochemistry parameterization together with a significant upward revision in the rate constant for aerobic methanotrophy. Nevertheless, our results strongly reinforce the arguments presented in Olson et al. (2016), and taken at face value further marginalize the role of $CH_4$ as a significant climate regulator at steady state during most of the Proterozoic Eon (between ~2.5 and 0.5 Ga).

Atmospheric $CH_4$ abundance is also strongly sensitive to the marine $SO_4^{2-}$ inventory (**Fig. 10b**). The scaling we observe between initial $SO_4^{2-}$ inventory and steady state atmospheric $pCH_4$ is very similar to that reported by Olson et al. (2016), with a sharp drop in the marine $CH_4$ inventory and atmospheric $CH_4$ abundance as marine $SO_4^{2-}$ drops below ~100 μmol kg$^{-1}$ (**Fig. 10b**). The implication is that for most of Earth history anaerobic oxidation of $CH_4$ in the ocean interior has served as an important inhibitor of $CH_4$ fluxes from the ocean biosphere. However, during much of the Archean Eon (between 4.0 and 2.5 Ga), sulfur isotope analysis indicates that marine $SO_4^{2-}$ concentrations may instead have been on the order of ~1-10 µmol kg$^{-1}$ (Crowe et al., 2014), while atmospheric $p$O$_2$ would also have been much lower than the values examined here (Pavlov & Kasting, 2002). The impact of the ocean biosphere and redox chemistry on atmospheric $p$CH$_4$ and Earth's climate system may thus have been much more important prior to ~2.5 billion years ago.

Interestingly, atmospheric $CH_4$ is significantly impacted by the value chosen for the biological energy quantum (BEQ). With all other parameters held constant, we observe an increase in steady state atmospheric $p$CH$_4$ from ~7 ppmv to ~25 ppmv when increasing the BEQ value from 20 to 30 kJ mol$^{-1}$ (**Fig. 10c**). This effect is mediated primarily by the importance of anaerobic methanotrophy when atmospheric $p$O$_2$ is low and the ocean interior is pervasively reducing. The standard free energy of AOM is of the same order of magnitude as the BEQ (see above), which elevates the importance of thermodynamic drive in controlling global rates of AOM. We would expect this effect to be much less important when aerobic methanotrophy is the predominant $CH_4$ consuming process within the ocean biosphere, as the standard free energy of this metabolism is over an order of magnitude greater than typical BEQ values for microbial metabolism (e.g., Hoehler, 2004). In any case, our results suggest that the role of thermodynamics should be borne in mind in scenarios for which AOM is an important process in the $CH_4$ cycle and seawater [$SO_4^{2-}$] is relatively low.

## 6. Discussion and Conclusions

The global biogeochemical cycling of $CH_4$ is central to the climate and redox state of planetary surface environments, and responds to the internal dynamics of other major biogeochemical cycles across a very wide range of spatial and temporal scales. There is thus strong impetus for the ongoing development of a spectrum of models designed to explore planetary $CH_4$ cycling, from simple box models to more computationally expensive 3-D models with dynamic and interactive ocean circulation. Our principal goal here is the development of a mechanistically realistic but simple and flexible representation of $CH_4$ biogeochemical cycling in Earth's ocean-atmosphere system, with the hope that this can be further developed to explore steady state and time-dependent changes to global $CH_4$ cycle in Earth's past and future and ultimately to constrain $CH_4$ cycling dynamics on Earth-like planets beyond our solar system.

To accomplish this, we have refined the organic carbon remineralization scheme in the cGENIE Earth system model to reflect the impact of anaerobic organic matter recycling in sinking aggregates within oxygenated waters, and to include the carbon cycling and isotopic effects of microbial $CH_4$ production. We have also incorporated revised schemes for microbial $CH_4$ consumption that include both kinetic and thermodynamic constraints, and have updated the parameterized atmospheric $O_2$-$O_3$-$CH_4$ photochemistry to improve accuracy and for use across a wider range of atmospheric $pO_2$ values than that explored in previous work. Simulations of roughly modern (high-$O_2$) and Proterozoic (low-$O_2$) Earth system states demonstrate that the model effectively reproduces the first-order features of the modern ocean-atmosphere $CH_4$ cycle, and can be effectively implemented across a wide range of atmospheric $O_2$ partial pressures and marine $SO_4^{2-}$ concentrations. In addition, our results strongly reinforce the conclusions of Olson et al. (2016) for the Proterozoic Earth system, while going beyond this to posit that the thermodynamics of anaerobic $CH_4$ consumption may have been important in regulating atmospheric $CH_4$ abundance during Archean time. Finally, our simulation of PETM-like carbon injection demonstrates the importance of explicitly considering $CH_4$ radiative forcing during transient warming events in Earth history.

We suggest that ongoing and future development work should focus on: (1) more rigorous tuning of organic carbon remineralization and $CH_4$ production/consumption schemes based on data fields from the modern ocean; (2) development and implementation of a more flexible parameterization of atmospheric photochemistry that allows the roles of atmospheric temperature structure, water vapor abundance, and atmospheric $pCO_2$ to be explored; (3) coupling of deep ocean chemistry with a description of marine methane hydrates and associated sedimentary $CH_4$ cycling; and (4) developing a representation of the production/consumption of $CH_4$ by terrestrial ecosystems.

## 7. Model code availability

A manual describing code installation, basic model configuration, and an extensive series of tutorials is provided. The Latex source of the manual and PDF file can be obtained by cloning (https://github.com/derpycode/muffindoc). The user manual contains instructions for obtaining, installing, and testing the code, as well as running experiments. The version of the code used in this paper is tagged as release v0.9.10 and has a DOI of 10.5281/zenodo.3620846. Configuration files for the specific experiments presented in the paper can be found in: cgenie.muffin/genie-userconfigs/MS/reinhardetal.GMD.2020. Details of the different experiments, plus the command line needed to run each, are given in README.txt.

**Author contributions:**

CTR, SLO, and AR developed new model code. CTR and CP compiled and analyzed empirical data for rates of methanotrophy. CTR performed all model simulations and data analysis. CTR prepared the manuscript with contributions from all co-authors.

**Acknowledgements:**

CTR acknowledges support from the NASA Astrobiology Institute (NAI), the Alfred P. Sloan Foundation, and the NASA Nexus for Exoplanet System Science (NExSS). SLO acknowledges support from the T.C. Chamberlin Postdoctoral Fellowship in the Department of Geophysical Sciences at the University of Chicago. AR, SKT, and YK were supported in part by an award from the Heising-Simons Foundation. We also thank Mark Claire for providing unpublished photochemical model results.

**TABLES:**

Table 1. Default parameters for organic matter production and water column remineralization.

| parameter | description | default value | units | source |
|---|---|---|---|---|
| *uptake/photosynthesis* | | | | |
| $\lambda$ | rate constant for DOM degradation | 0.5 | $y^{-1}$ | 1 |
| $\upsilon$ | fractional partitioning into DOM | 0.66 | — | 1 |
| $\tau_{bio}$ | nutrient uptake timescale | 63 | d | 2 |
| $\kappa_I$ | light limitation term | 20 | $W\ m^{-2}$ | 3 |
| $\kappa_P$ | half-saturation constant for $PO_4$ uptake | $2.1 \times 10^{-7}$ | $mol\ kg^{-1}$ | 1 |
| $k_{T0}$ | pre-exponential temperature constant | 0.59 | — | [see text] |
| $k_{eT}$ | exponential temperature constant | 15.8 | — | [see text] |
| *organic remineralization* | | | | |
| $r_1^{POM}$ | partitioning into labile POM fraction | 0.945 | — | 1 |
| $l_1^{POM}$ | *e*-folding depth for labile POM fraction | 589 | m | [see text] |
| $r_2^{POM}$ | partitioning into refractory POM fraction | 0.055 | — | 4 |
| $l_2^{POM}$ | *e*-folding depth for recalcitrant POM fraction | $10^6$ | m | [see text] |
| $\kappa_{O_2}$ | half-saturation constant for aerobic respiration | $1.0 \times 10^{-7}$ | $mol\ kg^{-1}$ | [see text] |
| $\kappa_{O_2}^i$ | inhibition constant for aerobic respiration | $1.0 \times 10^{-6}$ | $mol\ kg^{-1}$ | [see text] |
| $\kappa_{SO_4}$ | half-saturation constant for sulfate reduction | $5.0 \times 10^{-4}$ | $mol\ kg^{-1}$ | 4 |
| $\kappa_{SO_4}^i$ | Inhibition constant for sulfate reduction | $5.0 \times 10^{-4}$ | $mol\ kg^{-1}$ | 4 |

[1]Ridgwell et al. (2007); [2]Meyer et al. (2016); [3]Doney et al. (2006); [4]Olson et al. (2016)

**Table 2.** Default kinetic and thermodynamic parameters for oceanic methane cycling. Activity coefficients are estimated for $T = 25°C$ and $S = 35‰$.

| parameter | description | default value | units | source |
|---|---|---|---|---|
| *kinetic parameters* | | | | |
| $k_{AER}$ | rate constant for aerobic methanotrophy | 0.10 | y$^{-1}$ | [see text] |
| $\kappa_O^{AER}$ | half-saturation constant for O$_2$ | 2.0 x 10$^{-5}$ | mol kg$^{-1}$ | [see text] |
| $k_{AOM}$ | rate constant for AOM | 0.01 | y$^{-1}$ | [see text] |
| $\kappa_S^{AOM}$ | AOM half-saturation constant for SO$_4^{2-}$ | 5.0 x 10$^{-4}$ | mol kg$^{-1}$ | 1 |
| *thermodynamic parameters* | | | | |
| $\Delta G_{r,AER}^0$ | standard free energy yield of aerobic methanotrophy | -858.967 | kJ mol$^{-1}$ | 2 |
| $\Delta G_{r,AOM}^0$ | standard free energy yield of AOM | -33.242 | kJ mol$^{-1}$ | 2 |
| $\Delta G_{BQ,AER}$ | minimum free energy for aerobic methanotrophy | -15.0 | kJ mol$^{-1}$ | [see text] |
| $\Delta G_{BQ,AOM}$ | minimum free energy for AOM | -15.0 | kJ mol$^{-1}$ | 2-5 |
| $\gamma_{CH_4}$ | activity coefficient for dissolved CH$_4$ | 1.20 | — | 6-8 |
| $\gamma_{CO_2}$ | activity coefficient for aqueous CO$_2$ | 1.17 | — | 9 |
| $\gamma_{O_2}$ | activity coefficient for dissolved O$_2$ | 1.14 | — | 10 |
| $\gamma_{HCO_3^-}$ | activity coefficient for dissolved HCO$_3^-$ | 0.58 | — | 11, 12 |
| $\gamma_{HS^-}$ | activity coefficient for dissolved HS$^-$ | 0.75 | — | 13 |
| $\gamma_{SO_4^{2-}}$ | activity coefficient for dissolved SO$_4^{2-}$ | 0.10 | — | 11 |
| $R$ | gas constant | 8.2144 x 10$^{-3}$ | kJ K$^{-1}$ mol$^{-1}$ | |
| $\chi$ | stoichiometric number | 1.0 | — | 14 |
| *isotopic parameters* | | | | |
| $\varepsilon_{CH_4}$ | methanogenesis isotope effect | -35.0 | ‰ | [see text] |

[1]Olson et al. (2016); [2]Regnier et al. (2011); [3]Schink (1997); [4]Hoehler et al. (2001); [5]Hoehler (2004); [6]Stoessell and Byrne (1982); [7]Cramer (1984); [8]Duan et al. (1992); [9]Johnson (1982); [10]Clegg and Brimblecombe (1990); [11]Ulfsbo et al. (2015); [12]Berner (1965); [13]Helz et al. (2011); [14]Dale et al. (2008)

**Table 3.** Default constants and coefficients for $CH_4$ gas exchange. All default parameter values derived from Wanninkhof (1992). Schmidt number coefficients are for $S = 35$‰.

| parameter | description | default value |
|---|---|---|
| $a_1$ | Bunsen temperature coefficient 1 | -68.8862 |
| $a_2$ | Bunsen temperature coefficient 2 | 101.4956 |
| $a_3$ | Bunsen temperature coefficient 3 | 28.7314 |
| $b_1$ | Bunsen salinity coefficient 1 | -0.076146 |
| $b_2$ | Bunsen salinity coefficient 2 | 0.043970 |
| $b_3$ | Bunsen salinity coefficient 3 | -0.0068672 |
| $c_1$ | Schmidt temperature coefficient 1 | 2039.2 |
| $c_2$ | Schmidt temperature coefficient 2 | 120.31 |
| $c_3$ | Schmidt temperature coefficient 3 | 3.4209 |
| $c_4$ | Schmidt temperature coefficient 4 | 0.040437 |
| $k$ | Gas exchange constant | 0.31 |

**FIGURES:**

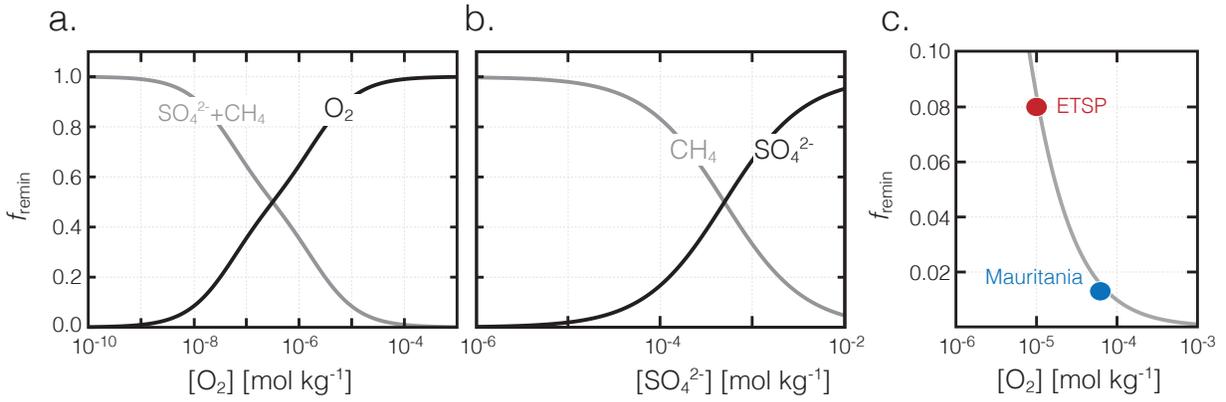

**Figure 1.** Fractional organic carbon remineralization by aerobic respiration, sulfate reduction, and methanogenesis in our modified organic matter remineralization scheme. In (a), relative rates of aerobic ($O_2$) and anaerobic ($SO_4^{2-}$ + $CH_4$) remineralization are plotted as a function of dissolved [$O_2$]. In (b), relative anaerobic remineralization rates are partitioned between sulfate reduction and methanogenesis as a function of dissolved [$SO_4^{2-}$] (dissolved [$O_2$] is held constant at $10^{-10}$ mol kg$^{-1}$). Shown in (c) are our estimated anaerobic remineralization fractions (grey curve) compared to estimates from a particle biogeochemical model applied to oxygen minimum zones (OMZs) in the Eastern Tropical South Pacific (ETSP) and Mauritanian upwelling (Bianchi et al., 2018).

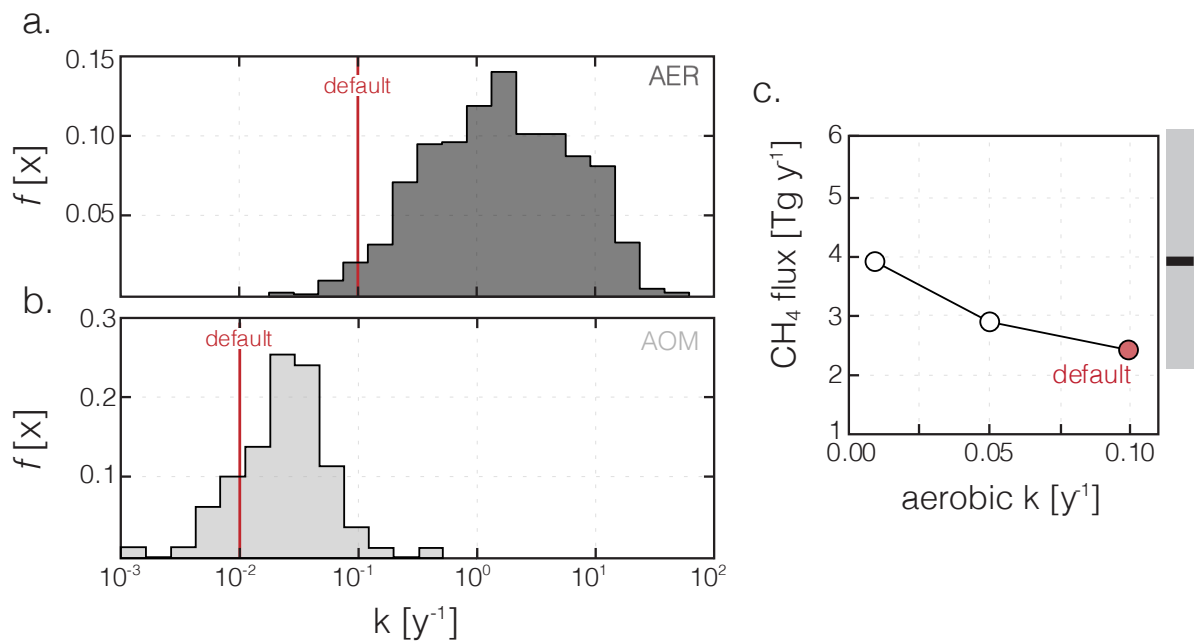

**Figure 2.** Compilation of rate constants for aerobic (AER; a) and anaerobic (AOM; b) methane oxidation. Rate constants are corrected for *in situ* temperature using a $Q_{10}$ of 2 (see Supplementary Materials). Vertical red lines show our default values as reported in **Table 2**. Shown in (c) are globally integrated diffusive fluxes of $CH_4$ from the ocean for a range of rate constants for aerobic methanotrophy, including our default simulation. The bar to the right of (c) shows the median (black bar) and 90% credible interval (grey shading) for estimates of the modern oceanic diffusive flux from (Weber et al., 2019)

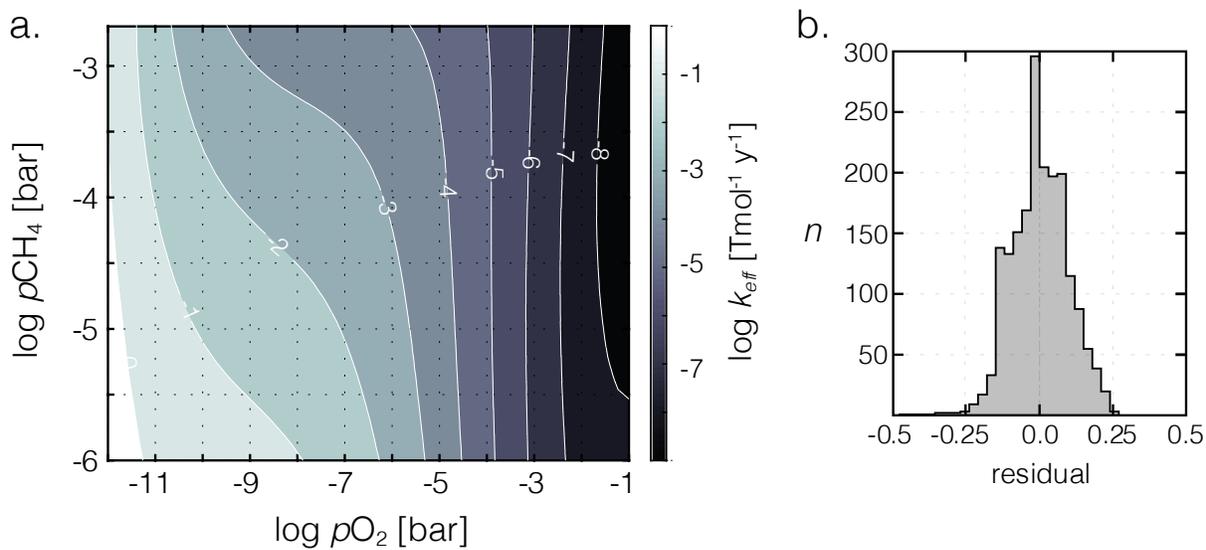

**Figure 3.** Shown in (a) is the bivariate fit to a suite of 1-D atmospheric photochemical runs for the effective rate constant ($k_{eff}$) parameterizing $O_2$-$O_3$-$CH_4$ photochemistry in ATCHEM. Shown in (b) is a frequency distribution of the residuals on $k_{eff}$ from the underlying photochemical model output.

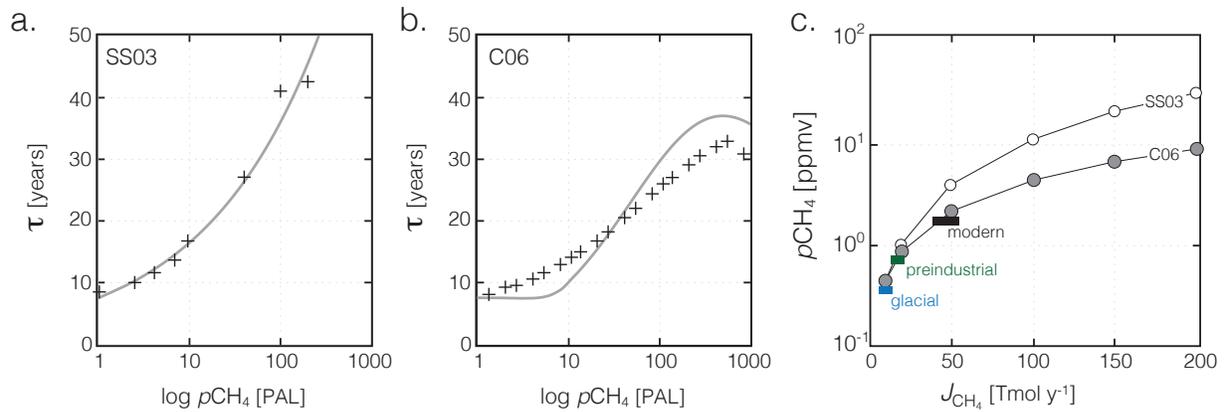

**Figure 4.** Comparison of steady-state atmospheric $p$CH$_4$ as a function of terrestrial CH$_4$ flux with modern/recent estimates. Shown in (a) is an exponential fit to the 2-D photochemistry model of Schmidt and Shindell (2003) (SS03), with individual model runs shown as black crosses. Shown in (b) is a plane through the bivariate fit shown in Figure 3 (grey curve), compared with the ensemble of 1-D atmospheric photochemical models at $p$O$_2$ = 0.1 atm (black crosses; see text). Shown in (c) are steady-state atmospheric CH$_4$ values as a function of imposed terrestrial CH$_4$ flux in our 'modern' configuration (circles), compared to estimates for the glacial, preindustrial, and modern CH$_4$ cycles (Kirschke et al., 2013;Bock et al., 2017;Paudel et al., 2016)

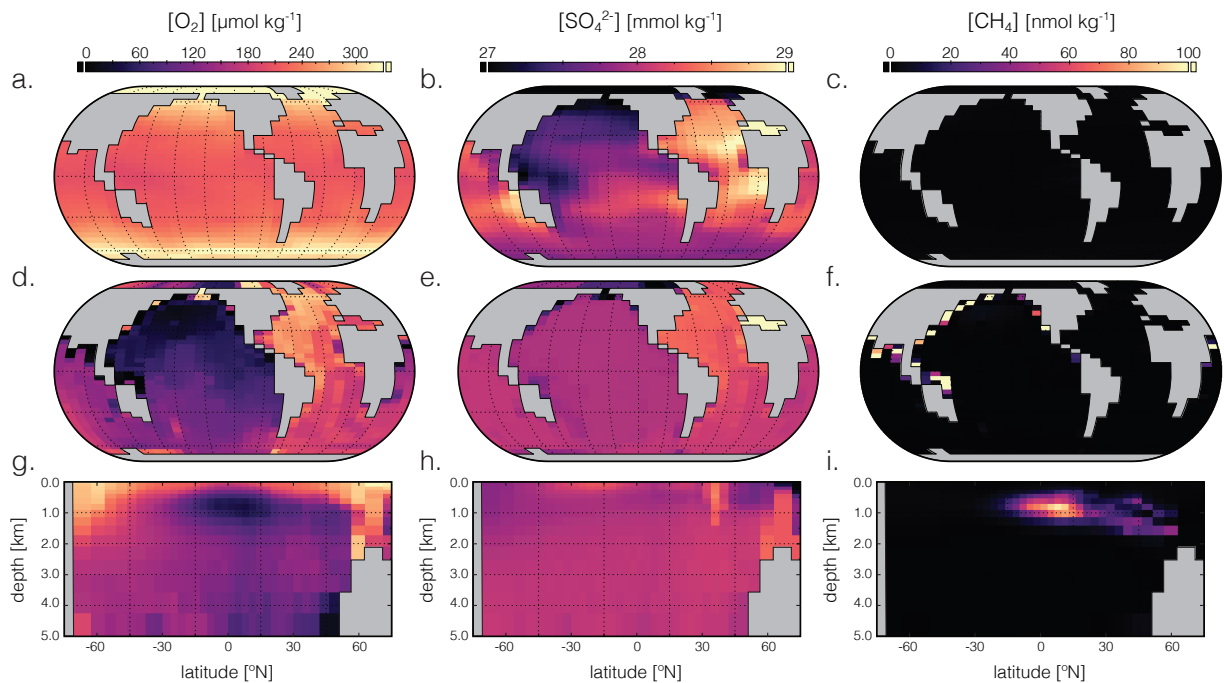

**Figure 5.** Tracer distributions in surface (a-c) and benthic (d-f) grid cells and in the zonally averaged ocean interior (g-i) for $O_2$ (a, d, g), $SO_4^{2-}$ (b, e, h), and $CH_4$ (c, f, i) in our 'modern' configuration. Note different concentration units for each tracer.

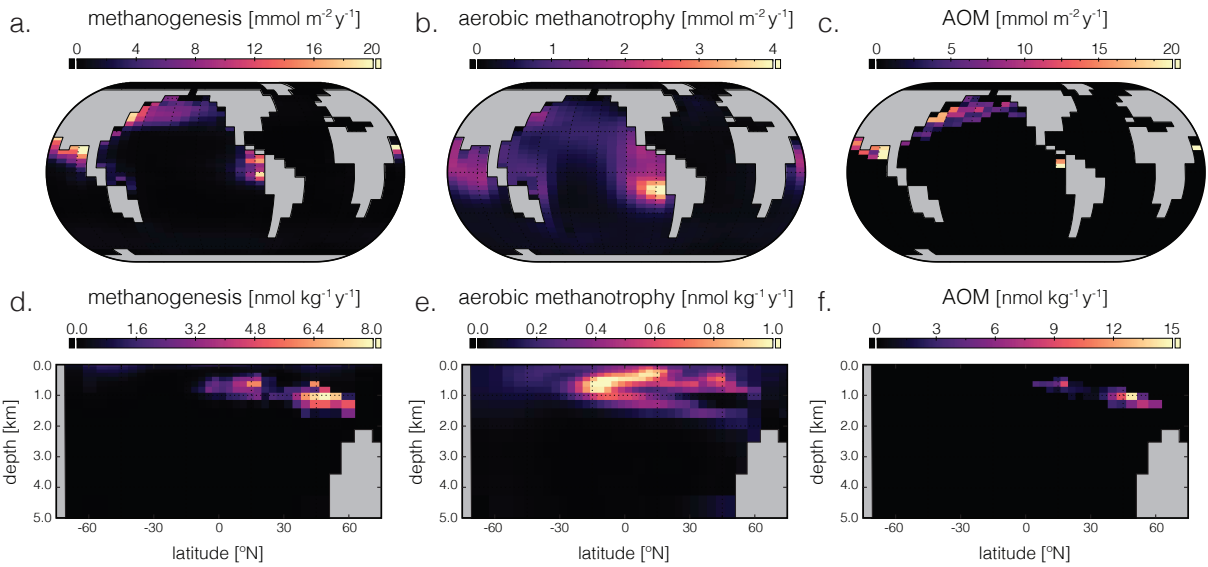

**Figure 6.** Major biological fluxes in the marine methane cycle for our 'modern' configuration. Panels show column integrated (a-c) and zonally averaged (d-f) rates of methanogenesis, aerobic methanotrophy, and anaerobic methane oxidation (AOM) in the ocean interior.

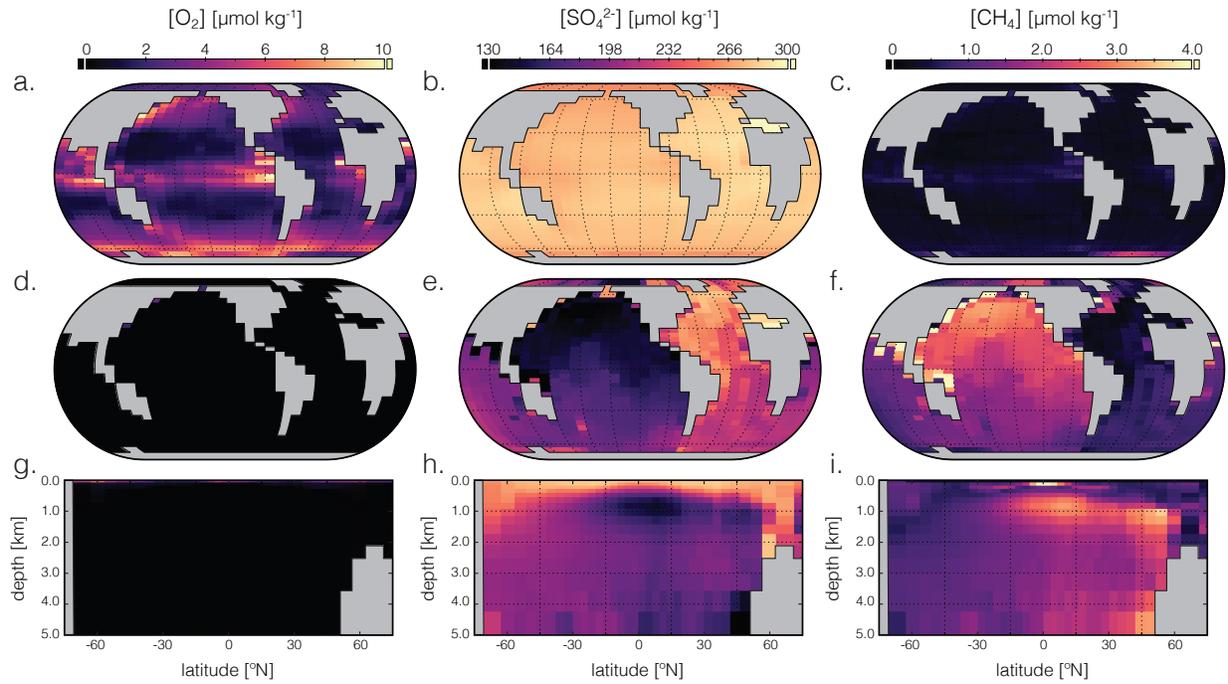

**Figure 7.** Tracer distributions in surface (a-c) and benthic (d-f) grid cells and in the zonally averaged ocean interior (g-i) for $O_2$ (a, d, g), $SO_4^{2-}$ (b, e, h), and $CH_4$ (c, f, i) in our 'ancient' configuration (see text). Note different concentration units for each tracer, and the differing scales relative to Figure 5.

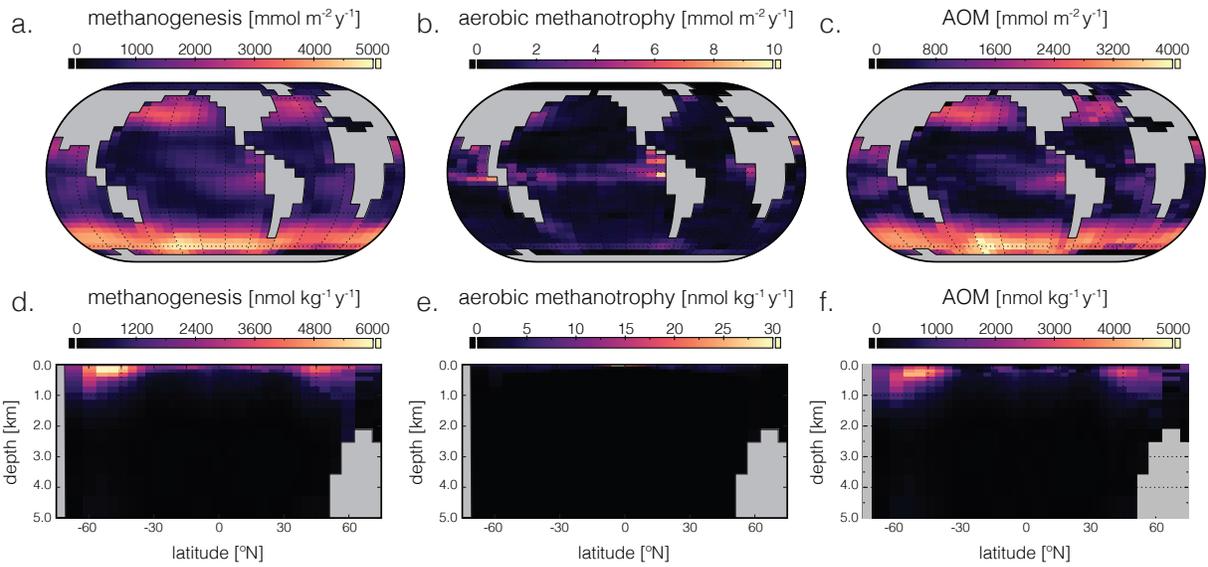

**Figure 8.** Major biological fluxes in the marine methane cycle for our 'ancient' configuration. Panels show column integrated (a-c) and zonally averaged (d-f) rates of methanogenesis, aerobic methanotrophy, and anaerobic methane oxidation (AOM) in the ocean interior.

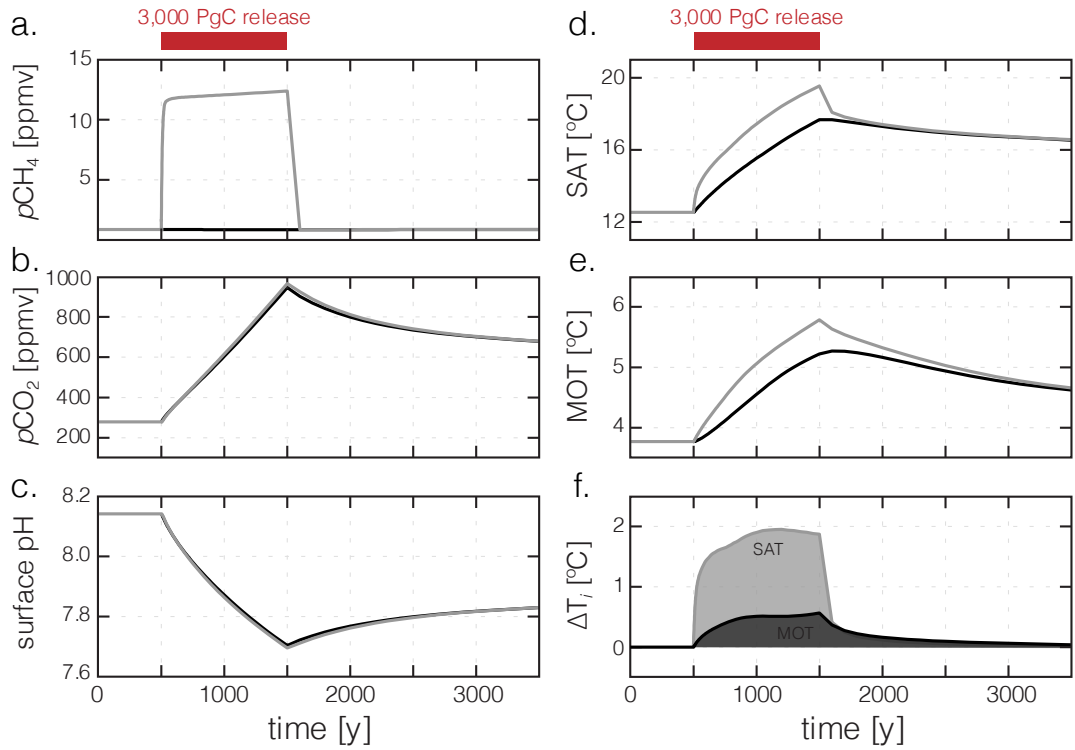

**Figure 9.** Response to a 3,000 PgC release directly to the atmosphere spread over 1,000 years, assuming carbon is injected as either $CH_4$ or $CO_2$. Atmospheric $pCH_4$ (a), $pCO_2$ (b), mean surface ocean pH (c), mean surface air temperature (SAT; d), and mean ocean temperature (MOT; e) are shown for a $CH_4$ injection (grey) and a $CO_2$ injection (black). Panel (f) shows the difference in SAT and MOT between the $CH_4$ and $CO_2$ injection scenarios ($\Delta T_i = T_{CH4,i} - T_{CO2,i}$) through time.

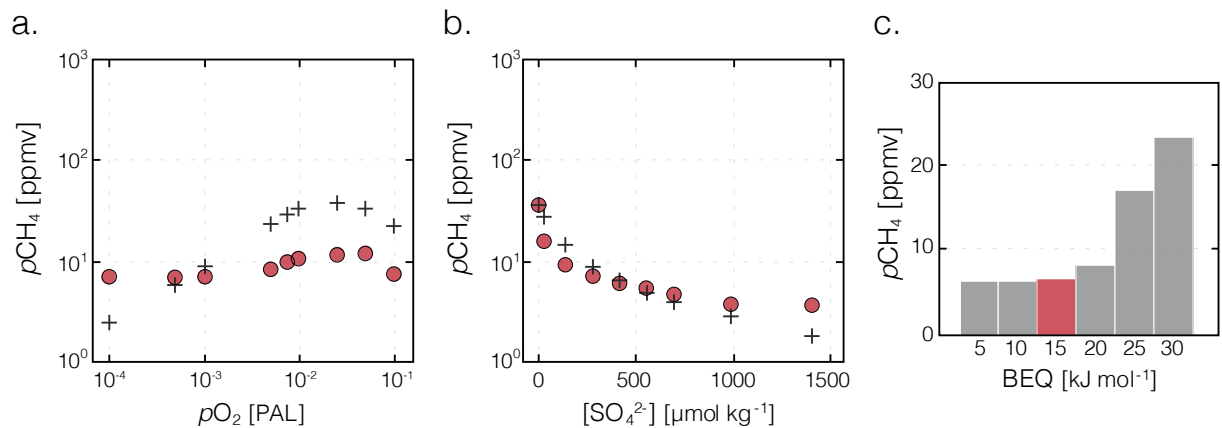

**Figure 10.** Sensitivity ensembles of our 'ancient' configuration compared to the results of Olson et al. (2016). Steady-state atmospheric $p$CH$_4$ values as a function of assumed atmospheric $p$O$_2$ (a) and initial marine SO$_4^{2-}$ inventory (b) are shown for our 'ancient' configuration (filled circles; see text) and from Olson et al. (black crosses). Shown below are additional ensembles showing the impact of varying the minimum free energy yield required for microbial methane oxidation (BEQ; c) on atmospheric $p$CH$_4$. All simulations were spun up from cold for 20 kyr, with the results shown from the last model year.